\documentclass{article}
\usepackage{amssymb,color}
\usepackage[notref,notcite]{showkeys}
\usepackage{epsfig}
\counterwithout{figure}{section}
\begin{document}
\def\E{\mathbb{E}}
\def\P{\mathbb{P}}
\def\R{\mathbb{R}}
\def\Z{\mathbb{Z}}
\def\scr{\scriptstyle}
\def\text{\textstyle}
\def\floor{\rm floor}
\def\sw{\rm sw}
\def\al{\alpha}
\def\be{\beta}
\def\ga{\gamma}
\def\om{\omega}
\def\Om{\Omega}
\def\12{\frac{1}{2}}
\def\oraw{\overrightarrow}
\def\nea{\nearrow}
\def\beq{\begin{equation}}
\def\eeq{\end{equation}}
\def\beqa{\begin{eqnarray}}
\def\eeqa{\end{eqnarray}}
\title{Composite states of wetting}
\author{Jo\"el De Coninck$^{\left( 1\right) }$,
 Fran\c cois Dunlop$^{\left( 2\right) }$ and
  Thierry Huillet$^{\left( 2\right) }$\\ \\
  $^{\left( 1\right) }$Laboratoire de Physique des Surfaces et Interfaces\\
Universit\'{e} de Mons, 20 Place du Parc, 7000 Mons, Belgium\\ \\
$^{\left( 2\right) }$Laboratoire de Physique Th\'{e}orique et Mod\'{e}lisation\\
CY Cergy Paris Universit\'e, CNRS UMR 8089,\\  
95302 Cergy-Pontoise, France\\
}

\maketitle
\begin{abstract}
The analytical expressions of liquid-vapor macroscopic contact angles are
analyzed for various simple geometries and arrangements of the substrate, in
particular when the latter exhibits two or more scales. It concerns the
Wenzel state of wetting when the substrate is completely wet, the
Cassie-Baxter state when the liquid hangs over the substrate, but also
intermediate states of wetting which are shown to be relevant and in
competition with the two other ones. Under a separation of scales
hypothesis, a composition rule of contact angles is developed whose interest
is illustrated in a close packing setup of rasperry-like particles.
\end{abstract}
\section{Introduction and outline of the results}
Superhydrophobicity refers usually to a surface which is sufficently rough to have a water droplet suspended on top of the roughness features, with air trapped underneath  corresponding thus to a water-air interface. Several approaches have been proposed to design such surfaces by creating  micro/nanostructures on hydrophobic surfaces. About the corresponding techniques let us mention crystallization control \cite{B13}, phase separation \cite{B14}, template synthesis \cite{B15,B16}, electrochemical deposition \cite{SMNCP}, and chemical vapor deposition \cite{B18}. Another way to proceed is to coat or graft micro/nanostructured hydrophilic surfaces with hydrophobic molecules \cite{B19,B20,B21}. Such materials include fluoroalkylsilanes \cite{B22}, fluoropolymers \cite{B23}, other organic polymers \cite{B24}, and alkylketene dimers \cite{B25}. To date, various initially hydrophilic substrates such as metal, glass, silicon wafer, and fabric have been used to prepare superhydrophobic surfaces in this way.

Among these techniques, one of the most popular method is to deposit one or two layers of nanoparticles on top of the base surface and to fix them by one way or another (sol-gel, annealing, grafting, ...). The advantage of this last technique is that it is cheap and simple to process.  Many examples can be found in the litterature \cite{McSN04,BGK05}. However what is not clear is what kind of nanoparticles should be used to enhance the superhydrophobic effect? This question has been addressed by many experimentalists where authors have considered several aspect ratios for one or two layers of nanoparticles \cite{TL07}. The purpose of the present work is to provide a thorough modelling of such complex wetting states.

Wetting issues have been addressed in a simplistic two-scale roughness
model in 1+1 dimensions \cite{DDH14}. Composite states of wetting are
already meaningful in this setup, together with the separation of
scales idea; see also \cite{DDH13} for a randomized setup and the robustness
problem of superhydrophobicity with respect to disorder.

In Section $2$, we discuss the problem of computing the Wenzel roughness $r$
for a $2$-scale substrate, as a function of the roughnesses $r_{1},r_{2}$ of
both scales separately. We find general conditions for $r=r_{1}r_{2}$ or for 
$r=r_{1}+r_{2}-1$. Then 
\[
\cos \theta ^{W}=r\cos \theta _{0},
\]
where $\theta ^{W}$ is the Young angle of a drop in the Wenzel (completely
wet) state and $\theta _{0}$ the Young angle for a flat surface of the same
material.

In Section $3$, we discuss the Cassie-Baxter covered fraction $\phi $ with
the liquid-vapor interface lying in a unique horizontal plane. With $r^{wet}$
the roughness of the wetted part of the substrate, we extend the validity of
the corresponding Young equation for the Cassie-Baxter contact angle: 
\[
\cos \theta ^{CB}=-1+\phi \left( 1+r^{wet}\cos \theta _{0}\right) .
\]

In Section $3.1$, we ask whether the liquid-vapour interface is stable,
metastable or unstable, following questions raised by Marmur \cite{Ma} who
showed stability for convex solids. His stability conclusions are extended to
include more general shapes that we call `vertically convex'.

In Section $3.2$, we investigate whether $\phi =\phi _{1}\phi _{2}$ for the
Cassie-Baxter wetting of $2$-scale substrates for which $r=r_{1}+r_{2}-1$
and with a roof-top structure at small scale on top of the one at the larger
scale. In this setup, under some restrictive hypothesis, composite states of
wetting can be defined in a natural way, together with their contact angles.

In Section $4$, the separation of scales hypothesis is introduced, allowing
to define a composition rule of contact angles. Subsection $4.1$ deals with
the case of a $2$-scale affine roughness situation, while Subsection $4.2$
focuses on the $3$-scale case. In Subsection $4.3$ a situation with
isotropic roughness is considered, still under the separation of scales
hypothesis. For some particular geometries, the expression of the
composition rule for contact angles is developed. This includes the case of
a close packing of rasperry-like particles.

As an input to the previous $2$-scale models, Section $5$ is devoted to $1$%
-scale models, with patterns obeying cylindrical symmetry. Subsection $5.1$
concerns monodisperse spherical particles. Under such hypothesis, the
contact angle $\theta ^{CB}$ of the Cassie-Baxter state is computed,
together with the contact angle $\theta ^{W}$ of the quasi-Wenzel state,
which has some air trapped at the basis of the spheres (small region bounded
by a catenoid). Subsection $5.2$ deals with monodisperse cylindrical
pillars, also possibly with air trapped below catenoids.

\section{Wenzel roughness}
The Wenzel roughness or specific surface area $r$ is defined as the true 
substrate area divided by the 
projected area. The definition applies to a surface or piece of a surface which
is plane at the macroscopic scale. The projection is onto the corresponding 
plane, so that the projected area is also simply the area on the macroscopic 
scale. The definition can be applied to any surface whose boundary is a plane
curve. The macroscopic surface is a plane domain $\Om\subset\R^2$, of area 
$|\Om|$ and piecewise smooth boundary $\partial\Om$. 
The real substrate is a surface $S$ embedded in $\R^3$ with the same boundary 
$\partial\Om$ as the macroscopic domain. The surface $S$ should be 
self-avoiding. We assume $S$ to be piecewise smooth.  Experimentally,
there is a scale length below which S is considered smooth. This scale is part
of the definition of the Wenzel roughness. We then define the Wenzel 
roughness as $r=\lim_{\Om\nearrow\R^2}\int_S dS/|\Om|$. Note that $|\Om|$ may be 
slightly different 
from the projected area of $S$, due to overhangs at the boundary, which should 
be negligible in the thermodynamic limit or may be absent altogether.
For $r$ to be uniquely defined, the sample should be large on a microscopic 
scale. In a microscopic model, one should take the limit as the area
tends to infinity, unless the real surface is periodic so that one cell is 
enough to compute. If the roughness 
is random, the existence of the limit requires a law of large numbers, which 
essentially requires independence between distant points on the substrate. 

As an example, partition the plane into squares of side $a$. In the center of 
each square grow a pillar of square section of fixed side $a_1<a$ and random
height $b_{1,i}$, where $i$ labels the pillar or the corresponding square.
The area is
\beq\label{S1}
S=\sum_i(a^2+4a_1b_{1,i})
\eeq
Assume the $b_{1,i}$ identically distributed, with mean $\bar b_1$.
The Wenzel roughness is
\beq\label{r1}
r_1=\lim_{\Om\nearrow\R^2}{S\over|\Om|}
=\lim_{\Om\nearrow\R^2}{\sum_i(a^2+4a_1b_{1,i})\over\sum_ia^2}
=1+4{a_1\bar b_1\over a^2}
\eeq
where the last step is the law of large numbers, easily satisfied if the
$b_{1,i}$ are asymptotically independent at large distances. 
A special case is $b_{1,i}$ taking some fixed value value $b_1$ with probability
$p$ and value 0 (no pillar) with probability $1-p$, so that $\bar b_1=pb_1$. 

If the real surface has (Hausdorff) dimension 2, not a fractal,
then the Wenzel roughness is scale invariant, as in (\ref{r1}). 
If the surface is defined by a single length, then the Wenzel roughness is 
independent of this length. For example a surface made by a compact
arrangement of monodisperse spheres upon a plane has a roughness independent
of the sphere radius.

In the completely Wenzel state the substrate is completely wet, with no air 
trapped at all. The corresponding contact angle $\theta^W$ is easily shown to 
obey 
\beq\label{wenzel}
\cos\theta^W=\lim_{\Om\nearrow\R^2}{S\,(\ga_{SV}-\ga_{SL})\over|\Om|\,\ga_{LV}}
=r\cos\theta_0\,,\qquad{\rm if}\ r\le 1/|\cos\theta_0|\,,
\eeq
 whatever the geometry 
\cite{W}. The angle $\theta_0$ is the Young angle for a flat surface of the same
material. 
If $r\cos\theta_0\ge1$, complete wetting occurs. If $r\cos\theta_0\le-1$ then the completely Wenzel state is not the minimizer of free energy. Indeed minimizing free energy is equivalent to maximizing the cosine of the apparent contact angle, and Cassie-Baxter configurations will have $\cos\theta^{CB}>-1$.

Measuring $r$ may be difficult. Parts of the real surface may be hidden by overhangs. The Wenzel roughness of superhydrophobic films built from raspberry-like particles \cite{Ming} is under-estimated by optical apparatus. 

In section \ref{spheres} we shall see almost Wenzel states, with a 
little air trapped below spherical particles, implying a deviation from Wenzel's
law (\ref{wenzel}).

\subsection{Two scales of roughness, $r=r_1r_2$ ?}\label{rr1r2}
There may be two or more mesoscopic scales of roughness, separated say by an 
order of magnitude. Here we examine the possibility that $r=r_1r_2$. At the 
first, larger, scale of roughness, one has a surface
$S_1$ embedded in $\R^3$ with the same boundary $\partial\Om$ as the 
macroscopic domain. The second, smaller, scale of roughness is not visible, so 
that $S_1$ is piecewise smooth and $\int_{S_1} dS_1=r_1|\Om|$. 
Now each $dS_1$ may be enlarged so that the second scale becomes visible,
and the true surface with the same boundary as $dS_1$ has an area $r_2dS_1$.
The true total area is $\int_{S_1} r_2dS_1=r_1r_2|\Om|$, the global roughness
is $r=r_1r_2$. The essential property here is that the roughness $r_2$ of $dS_1$
is independent of $dS_1$ and in particular independent of the orientation of 
$dS_1$ with respect to the macroscopic surface.

One can construct periodic surfaces with two scales not separated by an order
of magnitude, keeping the property that the roughness $r_2$ of $dS_1$
is independent of $dS_1$, where now $dS_1$ is not infinitesimal but is a face 
of a cell. The following construction is inspired by fractals.
\begin{figure}
\resizebox{\textwidth}{!}{\includegraphics{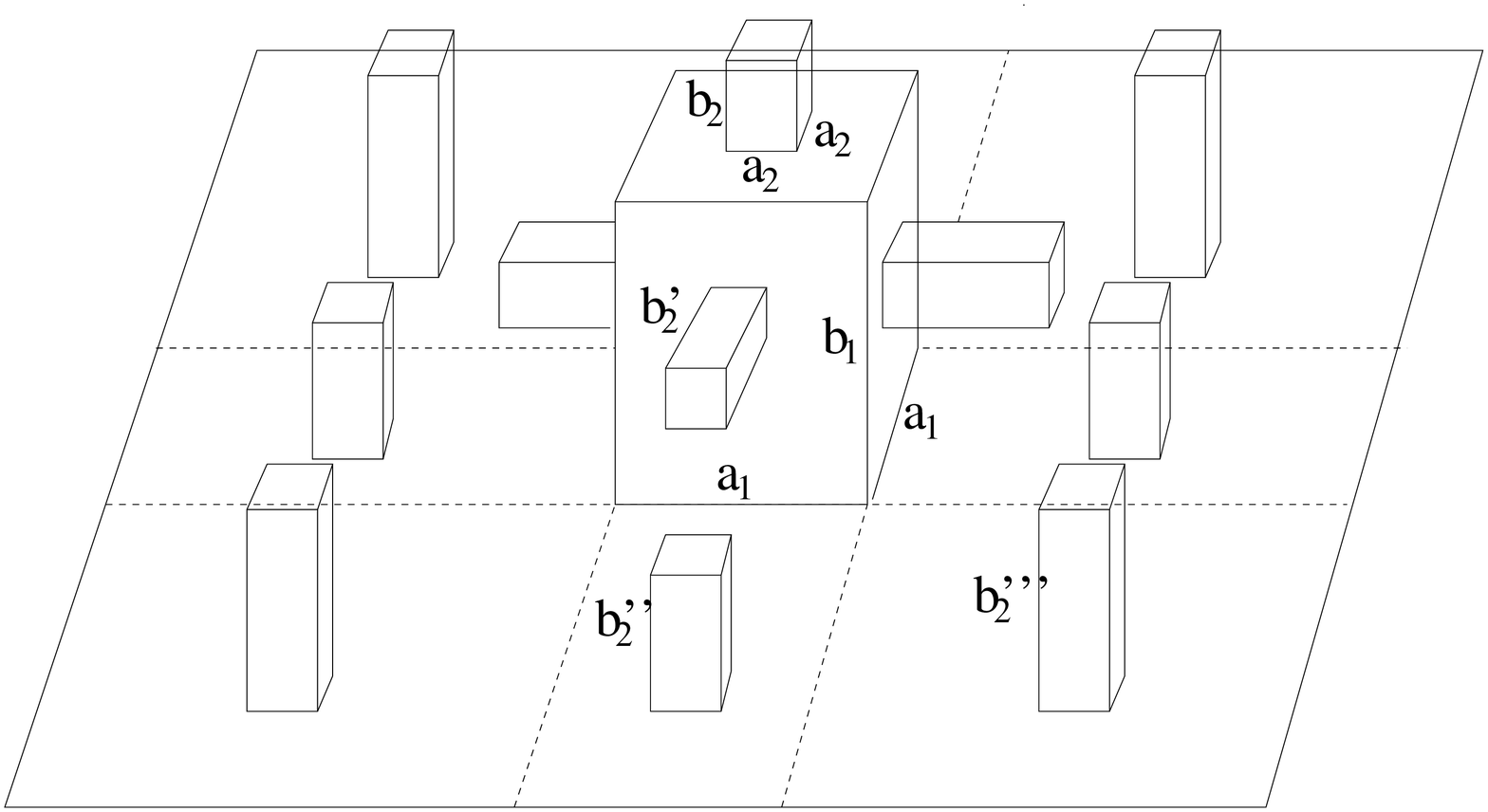}}
\caption{Cell of a periodic substrate with $r=r_1r_2$}\label{r1r2}
 \end{figure}
The elementary cell is a square of side $a$ (Fig. \ref{r1r2}). The first scale
of roughness is a pillar of square section of side $a_1$ and height $b_1$ placed
at the center. This gives a roughness $r_1=1+4a_1b_1/a^2$. The surface can now 
be seen as the union of 13 squares or rectangles: one square of side $a_1$,
4 squares of side $(a-a_1)/2$, 4 rectangles $a_1\times(a-a_1)/2$ and the 4
vertical rectangles $a_1\times b_1$. A second scale of roughness is introduced
by placing a pillar of square section of side $a_2$ and suitable height at the 
center of each of the 13 faces of the first scale. The height of the pillar 
will be in proportion of the area of the host face, which may be difficult in 
the laboratory but may be similar to some cases that occur in Nature. 
Here the square face of side $a_1$ will host a pillar of height $b_2$; 
the vertical rectangles $a_1\times b_1$ will host pillars of height 
$b'_2=b_2b_1/a_1$;
the rectangles $a_1\times(a-a_1)/2$ will host pillars of height 
$b''_2=b_2(a-a_1)/(2a_1)$; and the square faces of side $(a-a_1)/2$ will 
host pillars of height $b'''_2=b_2(a-a_1)^2/(4a_1^2)$. Now all 13 faces have a 
roughness $r_2=1+4a_2b_2/a_1^2$. Therefore the composite elementary cell and the
resulting periodic substrate have a roughness $r=r_1r_2$.

\begin{figure}
\begin{center}
\resizebox{9cm}{!}{\includegraphics{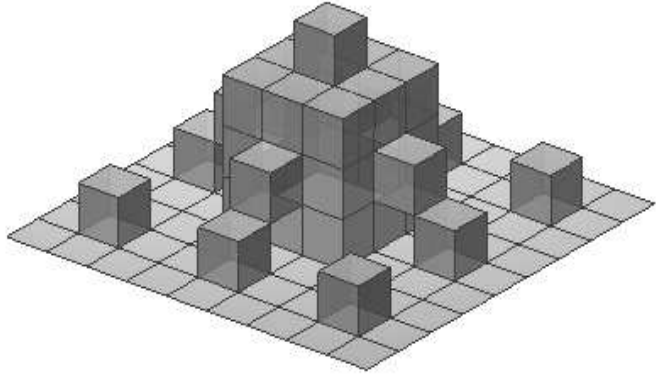}}
\caption{Quadratic Koch surface \cite{K2}}\label{koch2}
\end{center}
\end{figure}
A special case is $a_1=a/3,\, b_1=a_1,\,a_2=a_1/3,\,b_2=a_2$, giving 
$r_1=r_2=13/9$ (Fig. \ref{koch2}).
Iterating indefinitely produces the 3D quadratic Koch surface (type 1) 
\cite{K2}, of Hausdorff dimension $\log(13)/\log(3)\simeq2.3347$ \cite{Kd}, and
infinite Wenzel roughness. 
After $n$ iterations the true area is $(13/9)^na^2$. When applied to wetting
substrates, the number of iterations is limited by the definition of surface
tension: $(1/3)^na$ should be larger than a few nm to allow a continuum
description of the liquid. 

Disorder can be introduced in the quadratic Koch surface
by growing the pillars each time with probability $p$ independently, as after
(\ref{r1}), now leading to a Wenzel roughness $r_1=1+4p/9$ at the first scale 
and $r=r_1r_2=(1+4p/9)^2$ with two scales. The crucial property to obtain
the product of roughnesses is that the random numbers of different scales be 
independent, because they come up multiplied. On the other hand the random 
numbers attached to different faces of the same scale are chosen identically 
distributed but not necessarily independent: indeed they come up together
in a sum to which we want to apply the law of large numbers. So at the same 
scale only asymptotic independence at large distances is required. In the 
example, faces belonging to the same pillar are dependent,
and in fact share a common random number, but faces belonging to different 
pillars are independent.
 
In order to obtain hydrophobic substrates, one may prefer $a_1<a/3$ and 
$b_1>a_1$, and then
$a_2, b_2$ taking care to avoid intersection of pillars. A sufficient condition
is $a_1+4b_1b_2/a_1+2a_2<a$. An example is 
$a_1=a/5,\,b_1=2a_1,\,a_2=a_1/5,\,b_2=2a_2$, with $r_1=r_2=33/25$.

Some disorder can be introduced:
suppose the height of each pillar is multiplied by an independent copy
of a random variable $\xi$ of mean 1 and sufficiently narrow support around 1.
Let $S^{(1,2)}$ denote the 2-scale random surface (and its area) over the 
elementary square $a\times a$; let $S^{(2)}_i$ for $i=1\dots13$ denote the 
random surface over the $i$-th square or rectangle $S^{(1)}_i$ constructed on 
the first scale. Then
\beq
S^{(1,2)}=\sum_{i=1}^{13}S^{(2)}_i
=\sum_{i=1}^{13}\Bigl(S^{(1)}_i+4a_2b_{2,i}(\xi^{(1)})\,\xi_i^{(2)}\Bigr)
\eeq
where $b_{2,i}$ is one of $b_2,\,b'_2(\xi^{(1)}),\,b''_2,\,b'''_2$ where 
$b'_2(\xi^{(1)})=b_2b_1\xi^{(1)}/a_1$, and $\xi^{(1)}$ and the $\xi_i^{(2)}$'s are
independent copies of $\xi$. The average over $\xi^{(2)}$ reads
\beq
\E^{(2)}S^{(1,2)}=\sum_{i=1}^{13}\Bigl(S^{(1)}_i+4a_2b_{2,i}(\xi^{(1)})\Bigr)
=\sum_{i=1}^{13}r_2S^{(1)}_i
\eeq
where $r_2=1+4a_2b_2/a_1^2$ and 
\beq
\E^{(1)}\E^{(2)}S^{(1,2)}=r_2\sum_{i=1}^{13}\E^{(1)}S^{(1)}_i=r_1r_2a^2.
\eeq
where $r_1=1+4a_1b_1/a^2$. Since $b_1$ and $b_2$ are now the average heights,
the quantities $r_1$ and $r_2$ are the one-scale roughnesses as in (\ref{r1}),
and the 2-scale substrate has roughness $r=r_1r_2$.

The law $r=r_1r_2$ is linked to scale invariance and isotropy, like fractals,
as opposed to urban landscape with terrasses playing a different role than the
walls, which may be self-affine but not fully self-similar. We have put 
emphasis on block geometries for comparison with the following, but one or two
iterations of the triangular Koch surface works as well as a guide to define
a 2-scale surface of Wenzel roughness $r=r_1r_2$.  

\subsection{Two scales of roughness, $r=r_1+r_2-1$ ?}\label{r1r21} 

Let us now consider surfaces made of vertical and horizontal pieces only,
$S=S_{\rm horiz}\cup S_{\rm vert}$. We assume also that the surface is given by
a height function like in columnar Solid-On-Solid models: overhangs are 
forbidden. The height function is piecewise constant, and equals zero at the 
boundary. The jumps define walls, making up $S_{\rm vert}$.
The macroscopic reference surface is plane, as before, and by convention 
horizontal. Then 
\beq
r={|S|\over |S_{\rm horiz}|}=1+{|S_{\rm vert}|\over |S_{\rm horiz}|}
\eeq
The formula can be applied to one-scale roughness, as in (\ref{S1})(\ref{r1}), 
with the same result. On the other hand,
two-scale roughness will be different: the vertical parts of the first
scale cannot be made rough at the smaller scale because this would produce 
overhangs. Vertical parts can be added, but the total horizontal area always
remains equal to the reference area. Therefore
\beqa
S^{(1)}&=&S_{\rm horiz}^{(1)}\cup S_{\rm vert}^{(1)}\cr
S^{(1,2)}&=&\bigl(S_{\rm horiz}^{(2)}\cup S_{\rm vert}^{(2)}\bigr)
\cup S_{\rm vert}^{(1)}\cr
r&=&{|S_{\rm horiz}|+ |S_{\rm vert}^{(2)}|+|S_{\rm vert}^{(1)}|\over|S_{\rm horiz}|}
=1+(r_1-1)+(r_2-1)\label{r1r2m1}
\eeqa
Where we have used $|S_{\rm horiz}^{(2)}|=|S_{\rm horiz}^{(1)}|=|S_{\rm horiz}|$.
Formula (\ref{r1r2m1}) generalizes easily, for n scales of roughness, to
\beq
r=1+\sum_{i=1}^n(r_i-1)
\eeq

Embedding or separating different scales of roughness may be tricky when walls
of different scales stand one above the other. This question has been addressed
in \cite{DM} in the context of an SOS model of wetting seen as a 
cylinders model: any SOS surface with zero boundary condition can be uniquely
defined by families of nested cylinders, positive (building up pillars) or
negative (digging). The essential condition given in \cite{DM}, in addition to
nestedness, is that the intersection between the vertical parts of any two
cylinders must be empty or reduced to a line or a point, not a surface of 
positive area: a square pillar of the second scale cannot be placed at ground
level against a pillar of the first scale, because it would hide part of the
vertical exposed solid area of the first scale. Indeed this would spoil 
(\ref{r1r2m1}). 

Example 1: Consider the model on Fig. 1 without the $b'_2$ overhangs. The unit
cell has one $b_1$-pillar and 9 $b_2$- or $b''_2$- or $b'''_2$-pillars. 
The argument above can be applied, with $r_1$ and $r_2$ as before, except that
the 4 vertical faces of the $b_1$-pillar have no roughness. Now we have 
$r=r_1+r_2-1<r_1r_2$.

\begin{figure}
\begin{center}
\resizebox{9cm}{!}{\includegraphics{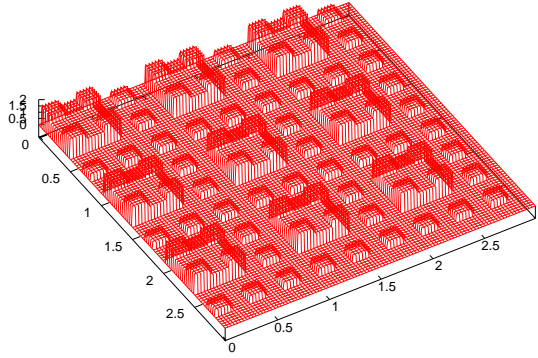}}
\caption{$r=r_1+r_2-1$}\label{sw}
\end{center}
\end{figure}
Example 2: Let $\sw(x)=2*\floor(x)-\floor(2*x)+1$ denote a square wave of 
wavelength 1 and amplitude 1, and $\sw(x,y)=\sw(x)\,\sw(y)$. A 2-scale profile 
\beq\label{sw12}
z(x,y)=b_1\,\sw(x/a_1,y/a_1)+b_2\,\sw(x/a_2,y/a_2)
\eeq
with $a_2=a_1/3$ and $b_2=b_1/2$ is shown on Fig. \ref{sw}. Its roughness is
\beq
r=r_1+r_2-1=1+{2b_1\over a_1}+{2b_2\over a_2}
\eeq
Here $a_1$ and $a_2$ are the periods, the respective pillars have side $a_1/2$
and $a_2/2$.

\section{Cassie-Baxter covered fraction}

If the substrate is not completely wet, there is a water-air interface. 
The pressure difference across the interface is neglected, 
a good approximation except in cases such as last stages of evaporation,
which we don't consider.
The interface is then a minimal surface, but not always a plane surface.
It may be disconnected into many pieces with different orientations,
due to air trapped in anfractuosities. 
\beqa\label{FSLCB0}
F_{SL}&=&\ga_{SL}A_{SL}+\ga_{SV}(A-A_{SL})+\ga_{LV}A_{LV}\cr
F_{SV}&=&\ga_{SV}A\cr
\cos\theta&=&\lim_{\Om\nearrow\R^2}{F_{SV}-F_{SL}\over|\Om|\,\ga_{LV}}
=\cos\theta_0\lim_{\Om\nearrow\R^2}{A_{SL}\over|\Om|}
-\lim_{\Om\nearrow\R^2}{A_{LV}\over|\Om|}
\eeqa

If the substrate is sufficiently regular, it may be composed of plane
horizontal pieces only, possibly at different heights on the substrate.
Or the liquid-vapor interface may even lie in a unique horizontal plane, whether
it be connected or disconnected. Such a configuration is often stable or 
metastable, with a maximum amount of air trapped, and we then call it a 
Cassie-Baxter state, in agreement with the simple examples.
This section is devoted to Cassie-Baxter states, and to other wetting states
with the liquid-vapor interface lying in a unique horizontal plane.

We first define the liquid-vapor interface fraction as 
the water-air interface area divided by the total projected area, 
$\phi_{LV}=|S_{LV}|/|\Om|$. The definition is straightforward because  $S_{LV}$ 
lies in a unique plane.

If the top level of the substrate corresponds to flat roof-tops all at the same
height, one may 
define a Cassie-Baxter state by placing the water-air interface at this top 
level. The covered (wetted) fraction is then $\phi=1-\phi_{LV}$ and is equal
to the top roof-tops area divided by total projected area.
The corresponding contact angle $\theta^{CB}$ is easily shown to obey \cite{CB}
\beq\label{CBflat}
\cos\theta^{CB}=(1-\phi)\cos\pi+\phi\cos\theta_0=-1+\phi\,(1+\cos\theta_0)
\eeq
whatever the geometry below the top level roof-tops, assuming some law of 
large numbers. If the top level of the substrate is rounded, like from coating
by spherical nanoparticles, the resulting formula (\ref{CBsph}) differs from
(\ref{CBflat}).
In all cases we define the covered fraction as $\phi=1-\phi_{LV}$, although it
may be smaller than the true wetted area divided by the total projected area.
The wetted part or parts of the solid are bounded by triple lines with the
air-water interface which is plane. Therefore this wetted part has a well 
defined Wenzel roughness which we denote $r^{\rm wet}$. Then the wetted area
is $S_{SL}=r^{\rm wet}\,\phi\,|\Om|$, and the solid-liquid and solid-vapor free
energies are
\beqa\label{FSLCB}
F_{SL}^{CB}&=&\ga_{SL}r^{\rm wet}\,\phi\,|\Om|
+\ga_{LV}(1-\phi)|\Om|+\ga_{SV}(r-r^{\rm wet}\,\phi\,)|\Om|\cr
F_{SV}&=&\ga_{SV}r|\Om|
\eeqa
leading with Young's equation to a macroscopic contact angle obeying
\beq\label{thetaCB}
\cos\theta^{CB}=-1+\phi\,(1+r^{\rm wet}\cos\theta_0)
\eeq
which coincides with formula (2.3) in \cite{BJK}. Note that
the roughness below the level of the air-water interface plays no role.
Formula (\ref{thetaCB}) will be applied to pillars of two 
scales in the next section, and to a substrate coated
with spherical nanoparticles in Section \ref{spheres}.

\subsection{Local stability from convexity}\label{convex}

When there is a liquid-vapour interface at a given height $z$, one may ask whether it is a stable or metastable or unstable configuration. The substrate roughness is assumed to be at a scale large compared to the molecular scale, so that a tangent plane to the solid is well defined except at edges and corners, and the angle between the liquid-vapour interface and the solid is well defined except at edges and corners. This angle must everywhere equal the Young angle $\theta_0$ for the configuration to be an extremum of the free energy. This local condition depends only upon the substrate geometry around height $z$ and is therefore unrelated to $r^{\rm wet}$ in (\ref{thetaCB}).

Let $A$ be a point on the solid surface and on the liquid-vapour interface, where they meet at an angle $\theta_0$. Consider the vertical plane containing $A$ and the normal vector to the solid surface at $A$. Let $B$ be a neighboring point in this plane on the solid surface. Consider moving the liquid-vapour interface from $A$ to $B$. Let $H$ be the projection of $B$ onto the horizontal plane at $A$. The change in free energy, per unit length perpendicular to the given vertical plane, is
\beq\label{conv1}
\pm AH\,\ga_{LV}\pm AB\,(\ga_{SV}-\ga_{SL})
\eeq
where the signs must be chosen according to whether the corresponding areas increase or decrease. The four cases may be handled one by one, and lead to the same conclusion. Consider the case where $B$ is higher than $A$ and the solid surface does not overhang, corresponding to $\theta_0\ge\pi/2$. Then (\ref{conv1}) is 
\beq
AH\,\ga_{LV}+AB\,(\ga_{SV}-\ga_{SL})=\ga_{LV}(AH+AB\,\cos\theta_0)
\eeq
Now $AH\simeq AB\,\cos(\pi-\theta_0)$ and $AH>AB\,\cos(\pi-\theta_0)$ if and only if the intersection of the solid with the given vertical plane is strictly convex in a neighborhood of $A$. Whence the conclusion:

A liquid-vapour interface at height $z$ is stable with respect to small fluctuations if and only if (i) it meets the solid at the same angle $\theta_0$ everywhere along the intersection, and (ii) at any point along the intersection, the intersection of the solid with the vertical plane containing the normal vector is locally strictly convex.

\begin{figure}
\begin{center}
\resizebox{9cm}{!}{\includegraphics{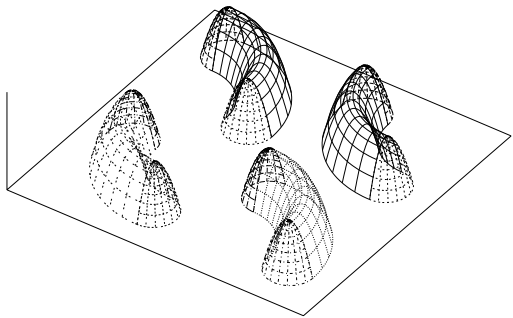}}
\caption{Vertically convex solids}\label{har}
\end{center}
\end{figure}

This extends \cite{Ma} so as to include non-trivial 3D geometries such as shown on Fig.\ref{har}, or hydrophobic legs of some insects.

Such solids can be constructed as follows. Let $\rho(z),\,z\in[0,z_{\max}]$ be a non-negative concave function, with $\rho(z_{\max})=0$. Let $\rho_{\max}=\max(\rho)$. Let $\cal C$ be a planar differentiable closed simple curve. 
For each $M\in\cal C$, let $M'=M+\rho_{\max}\vec n$ be the point at distance $\rho_{\max}$ from $M$ on the outer normal to $\cal C$ at $M$. Let $\cal C'$ be the set of such $M'$'s. Assume $d(M',\cal C)=\rho_{\max}$ for all $M'\in\cal C'$.
This is always true when the interior of $\cal C$ is a convex set, in particular if the curvature of $\cal C$ has a constant sign, positive for definiteness.
Otherwise, where the curvature is negative, the radius of curvature should be larger than or equal to $\rho_{\max}$ in order to satisfy the assumption at least locally.

Embed $\cal C$ and $\cal C'$ in the plane $z=0$ in three-dimensional space. At each point $M\in\cal C$, embed the graph $\rho(z)$ in the vertical plane containing $M$ and the normal vector to $\cal C$ at $M$, with $\{\rho=0\}$ as the vertical at $M$ and $\rho>0$ corresponding to the exterior of $\cal C$. Complete the construction of the solid with a roof at height $z_{\max}$ above the interior of $\cal C$. The projection of the solid onto the horizontal plane coincides with the interior of $\cal C'$.

We call such solids, and limits of such solids, ``vertically convex''. The roof, which is a translate of the interior of $\cal C$, may be shrinked to a line as in Fig.~\ref{har}, or to a point, giving a solid with cylindrical symmetry, such as a paraboloid of revolution or a sphere. A Cassie-Baxter interface on a periodic or random arrangement of identical vertically convex solids on a plane will be at the unique height $z(\theta_0)$, if any, where $\tan\theta_0=d\rho/dz$. Indeed the concavity of $\rho(z)$ implies that $d\rho/dz$ is monotone decreasing as a function of $z$. The Cassie-Baxter configuration will be locally stable, with (i) and (ii) above satisfied, if $\rho(z)$ is strictly concave at $z(\theta_0)$.   

Whether the Cassie-Baxter configuration is stable, being a global minimum of free energy, or metastable, being only a relative minimum, is a global question which can only be decided on a case by case basis. Of course a large $\theta_0$ favours stability.

Example: consider on a plane a close packing of identical spheres. If $\theta_0=\pi/2$, corresponding to $\ga_{SL}=\ga_{SV}$, the Cassie-Baxter interface is at the equator of the spheres. The excess in free energy of the Cassie-Baxter configuration relative to the Wenzel configuration comes solely from the small pieces of liquid vapour interface between the spheres, $(1-{\pi\over2\sqrt{3}})\,\ga_{LV}\simeq0.093\,\ga_{LV}$ per unit area. This is to be compared to the free energy barrier along a homogeneous path from Cassie-Baxter to Wenzel, $\ga_{LV}$ per unit area. The Cassie-Baxter configuration is metastable, protected by a significant free energy barrier. The resulting contact angle is $\cos\theta^{CB}=-1+{\pi\over2\sqrt{3}}$, giving $\theta^{CB}\simeq95^\circ$. The Wenzel configuration, with $\theta^W=90^\circ$, is the stable configuration.

The picture changes continuously when $\theta_0$ is varied in a neighborhood of $\pi/2$. When $\theta_0>\pi/2$, the stable ''Wenzel'' configuration becomes composite, with air trapped at the bottom of the spheres, and the corresponding $\theta^W$ as (\ref{Wsph}). When $\theta_0<\pi/2$, the metastable Cassie-Baxter configuration can give a slight hydrophobicity with a slightly hydrophilic material.

\subsection{Two scales of roughness, $\phi=\phi_1\phi_2$ ?}\label{p1p2}

Here we investigate whether $\phi=\phi_1\phi_2$ for Cassie-Baxter wetting of
2-scale substrates for which we have obtained the roughness $r=r_1+r_2-1$. 
We restrict further as follows: each of the two scales alone has a unique
height, $b_1$ for scale 1 and $b_2<b_1$ for scale 2, so that
\beqa
S_{\rm horiz}^{(1)}=S_{\rm horiz}^{(1,z=0)}\cup S_{\rm horiz}^{(1,z=b_1)},\quad
\phi_1=|S_{\rm horiz}^{(1,z=b_1)}|/|S_{\rm horiz}|\cr
S_{\rm horiz}^{(2)}=S_{\rm horiz}^{(2,z=0)}\cup S_{\rm horiz}^{(2,z=b_2)},\quad
\phi_2=|S_{\rm horiz}^{(2,z=b_2)}|/|S_{\rm horiz}|\cr
\eeqa
Applying roughness 2 on top of roughness 1 then gives
\beqa
S_{\rm horiz}^{(1,2)}=S_{\rm horiz}^{(1,2,z=0)}\cup S_{\rm horiz}^{(1,2,z=b_1)}
\cup S_{\rm horiz}^{(1,2,z=b_2)}\cup S_{\rm horiz}^{(1,2,z=b_1+b_2)}\cr\cr
S_{\rm horiz}^{(1,2,z=0)}=S_{\rm horiz}^{(1,z=0)}\cap S_{\rm horiz}^{(2,z=0)},\quad
S_{\rm horiz}^{(1,2,z=b_1)}=S_{\rm horiz}^{(1,z=b_1)}\cap S_{\rm horiz}^{(2,z=0)},\quad
\cr\cr
S_{\rm horiz}^{(1,2,z=b_2)}=S_{\rm horiz}^{(1,z=0)}\cap S_{\rm horiz}^{(2,z=b_2)},\quad
S_{\rm horiz}^{(1,2,z=b_1+b_2)}=S_{\rm horiz}^{(1,z=b_1)}\cap S_{\rm horiz}^{(2,z=b_2)}
\eeqa
The water-air interface in the true Cassie-Baxter state on the 2-scale
substrate will be at height $b_1+b_2$, and the corresponding covered fraction is
\beq
\phi={|S_{\rm horiz}^{(1,2,z=b_1+b_2)}|\over|S_{\rm horiz}|}
={|S_{\rm horiz}^{(1,z=b_1)}\cap S_{\rm horiz}^{(2,z=b_2)}|\over|S_{\rm horiz}|}
\eeq
The question now is whether the intersection is in proportion. If
\beq\label{hypfi12}
|S_{\rm horiz}^{(1,z=b_1)}\cap S_{\rm horiz}^{(2,z=b_2)}|=
{|S_{\rm horiz}^{(1,z=b_1)}|\over|S_{\rm horiz}|}|S_{\rm horiz}^{(2,z=b_2)}|
\eeq
then $\phi=\phi_1\phi_2$ follows. 
For example, if the two scales are well separated and the second roughness
is applied uniformly over all horizontal parts of the first scale, then
(\ref{hypfi12}) and $\phi=\phi_1\phi_2$ hold true to a good approximation. 
On the other hand, on a periodic substrate where the elementary cell has only
a few pillars, then (\ref{hypfi12}) may be wrong by a large error. This is
the case in the example of Fig. \ref{sw} corresponding to (\ref{sw12}). But the 
proportion can be respected: instead of (\ref{sw12}) with $a_2=a_1/3$, choose
\beq\label{sw128}
z(x,y)=b_1\,\sw(x/a_1,y/a_1)+b_2\,\sw(x/a_2-1/8,\,y/a_2-1/8)
\eeq
with $a_2=a_1/4$, still giving $r=r_1+r_2-1$ but now also $\phi=\phi_1\phi_2$
because the surfacic density of pillars at the second scale is the same on
the roof-tops of the first scale as on the pavement. 

On such a substrate there is a Wenzel state with the interface at height 0,
which we denote $W_{12}$; there is a Cassie-Baxter state at height $b_1+b_2$, 
which we denote $CB_{12}$; there is also a mixed state at height $b_1$, which we 
denote $CB_1W_2$ because the second scale of roughness on top of the first is 
completely wet. Then the usual formula for the Wenzel state, and (\ref{thetaCB})
applied to the other two states give
\beqa\label{CB1W2a}
\cos\theta^{W_{12}}=(r_1+r_2-1)\cos\theta_0\cr
\cos\theta^{CB_{12}}=-1+\phi_1\phi_2\,(1+\cos\theta_0)\cr
\cos\theta^{CB_1W_2}=-1+\phi_1\,(1+r_2\cos\theta_0)
\eeqa
It is worth noting that the $CB_1W_2$ configuration for substrate (\ref{sw128})
is stable or metastable, because the liquid-vapor interface is plane and is 
bounded by substrate edges and therefore locally minimal. On the contrary a
stable or metastable 
$CB_1W_2$ configuration for substrate (\ref{sw12}) cannot be planar
because the tentative plane meets vertical parts of substrate
at an angle $\pi/2$, in general different from the required $\theta_0$, away 
from edges (see Fig. \ref{sw}).


\section{Separation of scales}
\subsection{Two-scale affine roughness}
Consider a two-scale substrate obeying the hypotheses of Subsections 
(\ref{r1r21}) and (\ref{p1p2}), so that 
\beqa\label{CB12W12}
\cos\theta^{W_{12}}=r\cos\theta_0\,,\quad r=r_1+r_2-1\cr
\cos\theta^{CB_{12}}=-1+\phi\,(1+\cos\theta_0)\,,\quad\phi=\phi_1\phi_2
\eeqa
Instead of using the composite $r$ and $\phi$, one can write (\ref{CB12W12})
as a composition rule for contact angles:
\beqa
\cos\theta^{W_1X_2}=(r_1-1)\cos\theta_0+\cos\theta^{X_2}\label{W1X2}\\
\cos\theta^{CB_1X_2}=-(1-\phi_1)+\phi_1\cos\theta^{X_2}\label{CB1X2}
\eeqa
(\ref{W1X2})(\ref{CB1X2}) follows from (\ref{CB12W12}) with corresponding 
choices
$X=W$ or $X=CB$, but can also be derived directly and applied to any wetting 
state at scale 2 whenever the corresponding contact angle is known.
Indeed $\cos\theta$ is minus the specific excess solid-liquid free energy
relative to the solid-vapor free energy, in units of $\ga_{LV}$. And of course
$\cos\theta_0$ is the same for the smooth solid. Then the first term in 
(\ref{W1X2}) is easily identified with the vertical contribution, and the second
term with the horizontal contribution; the first term in (\ref{W1X2}) is the
liquid-vapor contribution, and the second term is the horizontal wetted top 
contribution.

Consider also the $CB_1W_2$ and $W_1CB_2$ configurations which are 
``Cassie-Baxter'' at scale 1 and ``Wenzel'' at scale 2 or conversely.
Separation of scales occurs when the thermodynamic limit 
is achieved for scale 2 roughness on each motif of scale 1 roughness.
In particular scale 2 motives at the boundaries of scale 1 motives are 
negligible with respect to the bulk. In this limit the 
curved parts of $CB_1W_2$ and $W_1CB_2$ liquid-vapor interfaces are negligible.
Indeed these curved parts border scale 2 motives at the boundaries of scale 1 
motives. Therefore (\ref{thetaCB})(\ref{CB1W2a}) apply,
\beq \label{CB1W2b}
\cos\theta^{CB_1W_2}=-1+\phi_1\,(1+r_2\cos\theta_0)
\eeq
And, using (\ref{FSLCB}) at scale 2, in terms of free energy densities
$f=\lim_{\Om\nea\R^2}F/|\Om|$,
\beqa
f_{SL}^{W_1CB_2}&=&\phi_2\ga_{SL}+(r_2-\phi_2)\ga_{SV}+(1-\phi_2)\ga_{LV}
+(r_1-1)\ga_{SL}\cr
f_{SV}&=&(r_1+r_2-1)\ga_{SV}
\eeqa
leading to
\beq\label{W1CB2}
\cos\theta^{W_1CB_2}=(r_1-1+\phi_2)\cos\theta_0-1+\phi_2
\eeq
Example \cite{SS}: $\theta_0=108^\circ$, 
scale 1 is a square array of discs of diameter $15\,\mu$m and height $2\,\mu$m,
with pitch $30\,\mu$m, leading to $r_1=1.1$. Scale 2 is a square array of discs 
of diameter $230\,$nm and height $500\,$nm, with pitch $430\,$nm, leading to
$r_2=2.95$, and $\phi_2=0.22$ in the $CB_2$ configuration. The combined 
roughness
is $r_1+r_2-1=3.05$, and (\ref{W1CB2}) gives $\theta^{W_1CB_2}=151^\circ$, winning
over the other wetting states, from (\ref{CB12W12})(\ref{CB1W2b}):
$\theta^{W_{12}}=161^\circ$, $\theta^{CB_{12}}=166^\circ$, $\theta^{CB_1W_2}=169^\circ$.

The authors of \cite{SS} find the same angle doing a computation in finite 
volume, and without assuming separation of scales. They do not however include
macroscopic boundary corrections associated with line tension nor scale 1
boundary corrections due partly to the fact that the larger pitch is not a 
multiple of the smaller, and partly due to the fact that the liquid vapor
interface must be curved on the boundary of scale 1 discs. They remark that the
value $151^\circ$ is less than the value $160^\circ$ found experimentally
\cite{SMNCP} for a substrate of similar roughness. The substrate in the
experiment, however, looks multi-scale rather than 2-scale.

\subsection{Three-scale affine roughness}
The method easily extends to three well separated scales of roughness. At each
scale wetting can be 'Cassie-Baxter' or 'Wenzel', hence $2^3=8$ states of 
wetting, with the following Young angles, the smallest of which is associated
with the stable configuration:
\beqa
1+\cos\theta^{CB_{123}}&=&\phi_3(1+\cos\theta^{CB_{12}})\\
1+\cos\theta^{CB_{12}W_3}&=&\phi_1\phi_2(1+\cos\theta^{W_{3}})\\
1+\cos\theta^{CB_1W_2CB_3}&=&\phi_1(1+\cos\theta^{W_2CB_3})\\
1+\cos\theta^{W_1CB_{23}}&=&\phi_2\phi_3(1+\cos\theta_0)+(r_1-1)\cos\theta_0\\
1+\cos\theta^{CB_1W_{23}}&=&\phi_1(1+\cos\theta^{W_{23}})\\
1+\cos\theta^{W_{12}CB_3}&=&\phi_3(1+\cos\theta_0)+(r_1+r_2-2)\cos\theta_0\\
1+\cos\theta^{W_1CB_2W_3}&=&\phi_2(1+\cos\theta_0)+(r_1-1+\phi_2(r_3-1))\cos\theta_0
\\
1+\cos\theta^{W_{123}}&=&1+(r_1+r_2+r_3-2)\cos\theta_0
\eeqa
\begin{figure}
\resizebox{\textwidth}{!}{\includegraphics{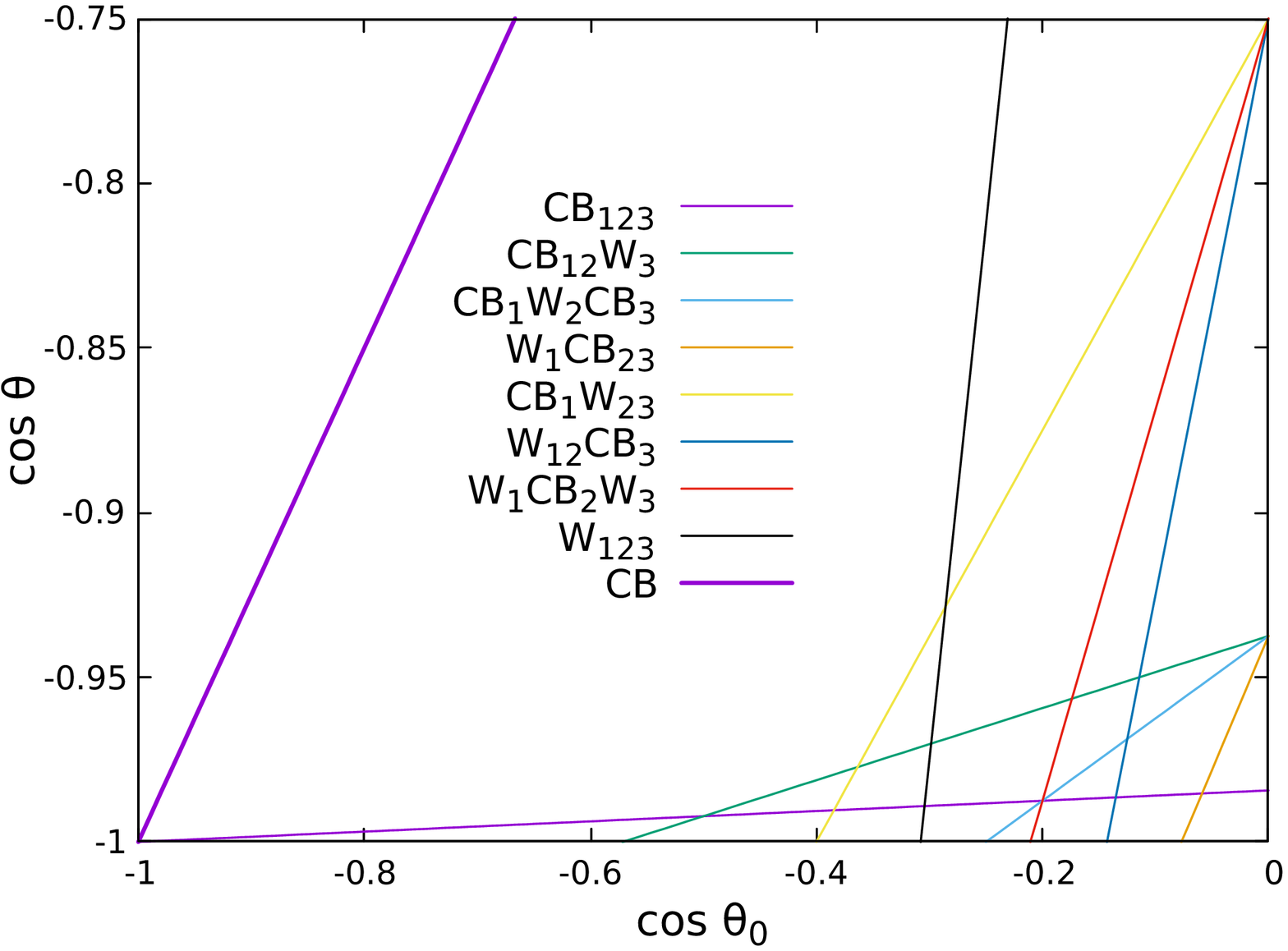}}
\caption{Eight wetting states with $r=r_1+r_2+r_3-2$, and 1-scale $CB$}\label{123}
 \end{figure}
Fig. \ref{123} shows $\cos\theta$ function of $\cos\theta_0$ for the eight
wetting states, together with a one-scale substrate of same Wenzel roughness.
The aspect ratios height/radius of cylinders making up the substrate on each
scale are the same. The roughness is $r_1=r_2=r_3=1.75$, the covered fractions
$\phi_1=\phi_2=\phi_3=0.25$.

\subsection{Isotropic roughness}
Consider a two-scale substrate obeying the main hypothesis of Subsection 
(\ref{rr1r2}): any surface element $dS_1$, infinitesimal at the larger scale,
has a true area $r_2dS_1$ when the smaller scale '2' is taken into account.
Suppose that the Young angle $\theta^{X_2}$ in a wetting state $X$ for scale 2 
roughness alone is known,
so that the corresponding free energy density is
\beq
f_{SL}^{X_2}=r_2\ga_{SV}-\cos\theta^{X_2}\ga_{LV}
\eeq
Then
\beq\label{W1X2I}
\cos\theta^{W_1X_2}={r_1r_2\ga_{SV}-r_1f_{SL}^{X_2}\over\ga_{LV}}=r_1\cos\theta^{X_2}
\eeq
giving
\beq
\cos\theta^{W_{12}}=r_1\cos\theta^{W_2}=r_1r_2\cos\theta_0
\eeq
Suppose in addition that $\theta^{CB_2}$ obeys (\ref{CBflat}). Then
\beq
\cos\theta^{W_1CB_2}=r_1\cos\theta^{CB_2}=-r_1+r_1\phi_2+r_1\phi_2\cos\theta_0
\eeq
Consider now the $CB_{12}$ and $CB_1W_2$ configurations, denoted $CB_1X_2$
with $X=CB$ or $X=W$:
\beq
f_{SL}^{CB_1X_2}=r_1^{\rm wet}\phi_1\,f_{SL}^{X_2}+(r-r^{\rm wet}\phi_1)\ga_{SV}
+(1-\phi_1)\ga_{LV}
\eeq
leading to
\beq\label{CB1X2I}
\cos\theta^{CB_1X_2}=-1+\phi_1(1+r_1^{\rm wet}\cos\theta^{X_2})
\eeq
Formulas (\ref{W1X2I})(\ref{CB1X2I}) agree with formulas previously derived
for simple geometries \cite{P}, and may be applied in more complex situations.

\smallskip\noindent{\sl Paraboloids:}

Consider at scale 1 a roughness made of paraboloids of revolution
$z=z_0-(x^2+y^2)/2a$, centered on an array or a random set of points
with density $\rho$, all rounded tops at the same $z_0$ with the same radius
of curvature $a$ at the top. In state $CB_1$, with the liquid-vapor interface 
meeting the paraboloids at an angle $\theta^{X_2}$, the covered fraction and
wet roughness are
\beq
\phi_1=\pi a^2(\tan\theta^{X_2})^2\rho\,,\qquad
r_1^{\rm wet}=-{\text {2\over3}}\, 
{1+(\cos\theta^{X_2})^3\over\cos\theta^{X_2}(\sin\theta^{X_2})^2}
\eeq
where we used the area of the paraboloid cap as
\beq
{\pi\over 6a^2}\Bigl(-{1\over(\cos\theta^{X_2})^3}-1\Bigr)
\eeq

\noindent{\sl Raspberries:}

A close packing of raspberry-like particles, with the smaller particles also close-packed, in the separation of scales limit, obeys (\ref{CBclose}) iterated once:
\beq\label{CB12close}
\cos\theta^{CB_{12}}=-1+{\pi^3\over24\sqrt{3}}(1+\cos\theta_0)^4
\eeq
giving $\theta^{CB_{12}}=144^\circ$ when $\theta_0=107^\circ$. 
If the smaller particles are not close-packed then $CB_1W_2$ may win.
For a density of the smaller particles equal to 0.5 divided by the area of an equilateral triangle of side $3R_2$, giving $\theta^{W_2}=138^\circ$ (see after (\ref{Wsph})), (\ref{CBclose}) with 138 instead of 107 gives $\theta^{CB_1W_2}=160^\circ$.

\section{Cylindrical symmetry}
Minimal surfaces with cylindrical symmetry are defined by ordinary differential equations, much easier mathematically than the more general partial differential equations. 
\subsection{Monodisperse spherical nanoparticles}\label{spheres}
Let us recall and reformulate in more generality some results for spherical 
nanoparticles \cite{DDH13}. The Wenzel state is not completely Wenzel when 
$\theta_0>\pi/2$ because some air is always trapped near the basis of the 
spheres. We don't have a general formula to take it into account, but the 
cylindrical symmetry allows an explicit computation of the corresponding
almost-Wenzel contact angle. And the top level is rounded, so that the 
Cassie-Baxter contact angle must be computed with (\ref{thetaCB}) instead of 
the usual simpler formula (\ref{CBflat}).

The random arrangement of nanoparticles makes a random triangulation of the 
plane, with the centers of the spheres as vertices of the triangulation.
Consider an area $A$ with $N$ nanoparticles, $A,N\to\infty$ with 
$4\pi R^2N/A\to r-1$. We assume that the disorder allows to apply the law of
large numbers, giving such a convergence and giving also the number of 
triangles as approximately twice the number of particles. Then, with $\sim$ 
from the law of large numbers,
\begin{eqnarray}\label{F3D}
&F_{SV}^{rr}(A)\sim(A+4\pi R^2N)\gamma_{SV} \hskip4cm\cr\cr
&F_{SL}^{CB}(A)\sim(A+2\pi(1-\cos\theta_0)R^2N)\gamma_{SV}+\hskip2.3cm\cr
&+2\pi(1+\cos\theta_0)R^2N\gamma_{SL}
+(A-\pi R^2\sin^2\theta_0N)\gamma_{LV} \cr\cr
&F_{SL}^{W}(A)\sim\Big(\pi\rho_0^2
+2\pi(R^2-R\sqrt{R^2-\rho_1^2})\Big)N\gamma_{SV}+\hskip1.3cm\cr
&+\Big(A-\pi\rho_0^2+2\pi(R^2+R\sqrt{R^2-\rho_1^2})\Big)\gamma_{SL}
+A_{\rm cat}N\gamma_{LV}
\end{eqnarray}
where $\rho_1$ and $\rho_0$ are the catenoid radii on the sphere and the plane 
respectively, and $A_{\rm cat}$ is the catenoid area between heights 0 and $z_1$,
\beq\label{catarea}
A_{\rm cat}=\pi\rho_c^2\Bigl({z_1\over\rho_c}+\12\sinh{2z_c\over\rho_c}
+\12\sinh{2z_1-2z_c\over\rho_c}\Bigr)
\eeq
In the Cassie-Baxter state, the covered fraction $\phi=1-\phi_{LV}$ and wetted
roughness $r^{\rm wet}$ are
\beqa
\phi={1\over4}(r-1)\sin^2\theta_0\cr
r^{\rm wet}={2(1+\cos\theta_0)\over\sin^2\theta_0}
\eeqa
so that applying (\ref{thetaCB}) gives
\beq\label{CBsph}
\cos\theta^{CB}=-1+{1\over4}(r-1)(1+\cos\theta_0)^2
\eeq
For close packing, $r=1+2\pi/\sqrt{3}$,
\beq\label{CBclose}
\cos\theta^{CB}=-1+{\pi\over2\sqrt{3}}(1+\cos\theta_0)^2
\eeq
giving $\theta^{CB}=123^\circ$ when $\theta_0=107^\circ$. Given $\theta_0$, close 
packing minimizes $\theta^{CB}$ and maximizes the stability of the $CB$ 
configuration.

The almost-Wenzel configuration is more complicated:
\begin{eqnarray}\label{F3D1}
&A^{-1}F_{SV}^{rr}(A)\sim r\,\gamma_{SV} \hskip5cm\cr\cr
&A^{-1}F_{SL}^{W}(A)\sim{1\over4}\Big(\hat\rho_0^2
+2(1-\sqrt{1-\hat\rho_1^2})\Big)(r-1)\gamma_{SV}
+\gamma_{SL}\hskip1.3cm\cr
&-{1\over4}\Big(\hat\rho_0^2-2(1-\sqrt{1-\hat\rho_1^2})\Big)(r-1)
\gamma_{SL}
+{\hat A_{\rm cat}\over4\pi}(r-1)\gamma_{LV}
\end{eqnarray}
where $\hat\rho_1$ and $\hat\rho_0$ and $\hat A_{\rm cat}$ should now be computed
for a sphere of radius one and are functions of $\theta_0$ only. Then
\beqa\label{Wsph}
\cos\theta^W=r\cos\theta_0
+{1\over4}\Big(\hat\rho_0^2+2(1-\sqrt{1-\hat\rho_1^2})\Big)\cos\theta_0\,(r-1)
-{\hat A_{\rm cat}\over4\pi}(r-1)\cr>r\,\cos\theta_0 \hskip8.5cm
\eeqa
For a density of nanoparticles equal to 0.5 divided by the area
of an equilateral triangle of side $3R$, giving $r=1+{8\pi\over9\sqrt{3}}$,
and $\theta_0=107^\circ$,
(\ref{Wsph}) gives $\theta^W=138^\circ$, winning over $\theta^{CB}=142^\circ$.

The Wenzel contact angle $\theta^W$ is a little smaller but very near the 
approximation $r\cos\theta_0$, at least when $\theta_0$ is not much bigger than
$\pi/2$. If either the Cassie-Baxter configuration or the Wenzel 
configuration minimize the energy in every triangle of the triangulation by
the nanoparticle centers, then the maximum between $\cos\theta^{CB}$ and 
$\cos\theta^W$ will be the true contact angle. This may be a good approximation
away from the value of $\cos\theta_0$ corresponding to the transition
between the two. However, except for periodic substrates, this transition will 
be smeared by the randomness.

\subsection{Monodisperse cylindrical pillars}
In the case of spherical nanoparticles, whatever $\pi/2<\theta_0<\pi$,
the liquid wets a nontrivial fraction of the sphere, fraction tending to zero
when $\theta_0\nearrow\pi$ and tending to one when $\theta_0\searrow\pi/2$. 
On the contrary, in the case of a pillar of finite height, depending upon 
$\theta_0$, the liquid may wet completely the pillar, or also de-wet completely 
the vertical surface of the pillar. The transitions can be 
computed exactly for pillars with cylindrical symmetry. 

Indeed the liquid-vapor interface is then again a catenoid as for spherical 
nanoparticles, $\rho=\rho_c\cosh\bigl((z-z_c)/\rho_c\bigr)$, where the catenoid
parameters $\rho_c,z_c$ have to be computed from the boundary conditions.
Gravity is neglected, as before. Consider the case where the 
triple line is at a height $z_1$ less than the height of the cylinder.
Let $\rho_0$ be the catenoid radius at the plane substrate, $z=0$. 
Then we have to solve
\beqa\label{cylcat}
\rho_0=\rho_c\cosh{z_c\over\rho_c}\,,\qquad
R=\rho_c\cosh{z_1-z_c\over\rho_c}\cr
\cot\theta_0=-\sinh{z_c\over\rho_c}\,,\qquad
\tan\theta_0=\sinh{z_1-z_c\over\rho_c}
\eeqa
which give, using $(\cosh x)^2-(\sinh x)^2=1$,
\beqa\label{cylsol}
\rho_c=-R\cos\theta_0\,,\qquad \rho_0=-R\cot\theta_0\hskip2cm\cr\cr
{z_c\over\rho_c}=-\sinh^{-1}(\cot\theta_0)\,,\qquad
{z_c-z_1\over\rho_c}=\cosh^{-1}\Bigl({-1\over\cos\theta_0}\Bigr)>0
\eeqa
which requires $\theta_0>3\pi/4$ as expected from the minimality of the surface 
and the boundary conditions. 
$\rho_0$ and $z_1$ vary respectively from $R$ to 
$+\infty$ and from 0 to $+\infty$ as $\theta_0$ varies from $3\pi/4$ to $\pi$.
When $z_1$ reaches the top of the pillar $z_1=b$, at some $\theta_0=\theta_0^b$,
the vertical surface of the pillar is completely de-wetted.
When $\theta_0>\theta_0^b$, the top level of the catenoid $z_1$ remains at 
$z_1=b$, but the contact angle
at  the top of the pillar becomes strictly less than $\theta_0$,
the last equation in (\ref{cylcat}) is lost. 
When $\pi/2<\theta_0<3\pi/4$, the liquid wets completely the pillar.
In summary, starting from $\theta_0=\pi/2$, the liquid wets completely the 
pillar until $\theta_0=3\pi/4$, then from $\theta_0=3\pi/4$ to 
$\theta_0=\theta_0^b$ the pillar gradually dewets, and then remains with a 
completely dewetted vertical surface up to  $\theta_0=\pi$.

For $\pi/2\le\theta_0\le 3\pi/4$ the Wenzel state is completely Wenzel, so that
$\cos\theta^W=r\cos\theta_0$. Let us now consider 
$3\pi/4\le\theta_0\le\theta_0^b$
and proceed, as for the spherical nanoparticles, from a triangle in the
triangulation by the cylinder centers on the plane substrate.
\beqa\label{FWcyl}
F_{SV}^{rr}=(A_{\rm tri}+\pi Rb)\gamma_{SV} \hskip5cm\cr\cr
F_{SL}^{W}=\Big({1\over2}\pi(\rho_0^2-R^2)+\pi Rz_1\Big)\gamma_{SV}\hskip4cm\cr
+\Big(A_{\rm tri}-{1\over2}\pi(\rho_0^2-R^2)+\pi R(b-z_1)\Big)\gamma_{SL}
+{1\over2}A_{\rm cat}\gamma_{LV}
\eeqa
Factors of 1/2 arise because the three angles in the triangle, cutting
cylinder and catenoid sectors, sum up to $\pi$ only. The full catenoid area
between heights 0 and $z_1$ is (\ref{catarea}) with (\ref{cylsol}),
\beq
A_{\rm cat}=\pi R^2\cos^2\theta_0\Bigl[-\cosh^{-1}\Bigl({-1\over\cos\theta_0}\Bigr)
-\sinh^{-1}(\cot\theta_0)-{\cos\theta_0\over\sin^2\theta_0}
-{\sin\theta_0\over\cos^2\theta_0}\Bigr]
\eeq
which is positive for $\theta_0>3\pi/4$.
From (\ref{FWcyl}) we get for this partially Wenzel state
\beqa
\cos\theta^W=r\cos\theta_0-{r-1\over 2\pi Rb}\,\Bigl[\,2\pi Rz_1\cos\theta_0
+\pi(\rho_0^2-R^2)\cos\theta_0+A_{\rm cat}\,\Bigr]\cr
>r\cos\theta_0 \hskip8cm
\eeqa
where $r=1+2\pi RbN/A$. For comparison, given the flat roof-tops,
\beq
\cos\theta^{CB}=-1+\phi^{CB}\,(1+\cos\theta_0)
\eeq 
with $\phi^{CB}=\pi R^2N/A$.

Let us now consider $\theta_0^b<\theta_0<3\pi/4$. The second and third
equations in (\ref{cylcat}) with $z_1=b$ give an equation for $\rho_c$ alone,
\beq
R=\rho_c\bigl(\cosh{b\over\rho_c}\bigr){1\over\sin\theta_0}
+\rho_c\bigl(\sinh{b\over\rho_c}\bigr)\cot\theta_0
\eeq
and then $\rho_0=\rho_c/\sin\theta_0$ and $z_c=\rho_c\,\cosh^{-1}(1/\sin\theta_0)$
and from (\ref{FWcyl}) with $z_1=b$, 
\beq
\cos\theta^W=\cos\theta_0-\pi(\rho_0^2-R^2)\cos\theta_0\,{N\over A}
-A_{\rm cat}\,{N\over A}
\eeq
where $N/A=(r-1)/(2\pi Rb)$ and $A_{\rm cat}$ is (\ref{catarea}) with $z_1=b$.

\medskip\noindent{\bf Acknowledgements}
The authors thank the European Space Agency (ESA) and the Belgian Federal
Science Policy (BELSPO) for their support in the framework of the PRODEX
Programme. This research was also partially funded by the  FNRS and the Region
Wallonne.

\end{document}